\documentclass[12pt,showpacs,preprintnumbers,amsmath,amssymb,
superscriptaddress,nofootinbib]{revtex4}

\newcommand{\tens}[1]{{\boldsymbol{#1}}}

\newcommand{\pa}{{\tens{\partial}}}

\newcommand{\be}{\begin{equation}}
\newcommand{\ee}{\end{equation}}
\newcommand{\ba}{\begin{eqnarray}}
\newcommand{\ea}{\end{eqnarray}}

\newcommand{\hh}{\, ,\hspace{0.5cm}}
\newcommand{\hhh}{\, ,\hspace{0.2cm}}

\newcommand{\hook}{\raisebox{-0.35ex}{\makebox[0.6em][r]
{\scriptsize $-$}}\hspace{-0.15em}\raisebox{0.25ex}{\makebox[0.4em][l]{\tiny
 $|$}}}
\newcommand{\eq}[1]{(\ref{#1})}

\newcommand{\n}[1]{\label{#1}}
\newcommand{\bss}[1]{{\boldsymbol{#1}}}

\newcommand{\bsn}[2]{{\boldsymbol{#1}}_{#2}}
\newcommand{\bsnn}[2]{{\boldsymbol{#1}}_{\bar{#2}}}

\newcommand{\ve}{\varepsilon}
\newcommand{\bse}[2]{{\boldsymbol{#1}}^{#2}}
\newcommand{\bsee}[2]{{\boldsymbol{#1}}^{\bar{#2}}}

\newcommand{\CQG}[3]{ #2 {\em Class. Quantum Grav.\ }{\bf #1} #3}
\newcommand{\JMP}[3]{ #2 {\em J. Math. Phys.\ }{\bf #1} #3}
\newcommand{\PRD}[3]{ #2 {{\em Phys. Rev.}\  D\ }{\bf #1} #3}
\newcommand{\CMP}[3]{ #2 {\em Comm. Math. Phys.}{\bf #1} #3}
\newcommand{\PLA}[3]{ #2 {{\em Phys. Lett.}\ A\ }{\bf #1} #3}
\newcommand{\PLB}[3]{ #2 {{\em Phys. Lett.}\ B\ }{\bf #1} #3}
\newcommand{\PRL}[3]{ #2 {\em Phys. Rev. Lett.\ }{\bf #1} #3}
\newcommand{\GRG}[3]{ #2 {\em Gen. Rel. Grav.\ }{\bf #1} #3}
\newcommand{\NC}[3]{ #2 {\em Nuovo Cimento\ }{\bf #1} #3}
\newcommand{\PRSA}[3]{ #2 {{\em Proc. R. Soc.}\ A\ }{\bf #1} #3}
\newcommand{\AM}[3]{ #2 {\em Ann. Math.} {\bf #1} #3}
\newcommand{\IJTP}[3]{ #2 {\em Int. J. Theor. Phys.\ }{\bf #1} #3}
\newcommand{\TEN}[3]{ #2 {\em Tensor\ }{\bf #1} #3}
\newcommand{\NPB}[3]{ #2 {{\em Nucl. Phys.}\ B\ }{\bf #1} #3}
\newcommand{\AP}[3]{ #2 {\em Ann. Phys., NY\ }{\bf #1} #3}
\newcommand{\IJMPA}[3]{ #2 {{\em Int.\ J.\ Mod.\ Phys.}\  A\ }{\bf #1} #3}
\newcommand{\PTP}[3]{ #2 {\em Prog. Theor. Phys.\ }{\bf #1} #3}

\begin{document}

\title{Higher-Dimensional  Black Holes: Hidden Symmetries
 and Separation of Variables}

\author{
Valeri P. Frolov
and David Kubiz\v n\'ak
}

\affiliation{
Theoretical Physics Institute, University of Alberta, Edmonton,
Alberta, Canada T6G 2G7
}

\email{frolov@phys.ualberta.ca, kubiznak@phys.ualberta.ca}

\begin{abstract}{ In this paper, we discuss hidden symmetries in
rotating black hole spacetimes. We start with an extended
introduction which mainly summarizes results on hidden symmetries in
four dimensions and introduces Killing and Killing--Yano tensors,
objects responsible for hidden symmetries. We also demonstrate how
starting with a principal CKY tensor  (that is a closed
non-degenerate conformal  Killing--Yano 2-form) in 4D flat
spacetime one can `generate' 4D Kerr-NUT-(A)dS solution and its
hidden symmetries. After this we consider higher-dimensional
Kerr-NUT-(A)dS metrics and demonstrate that they possess a principal
CKY tensor which  allows one to generate the whole  tower of
Killing--Yano and Killing tensors. These symmetries imply complete
integrability of geodesic equations and complete separation of
variables for the Hamilton--Jacobi, Klein--Gordon,  and Dirac
equations in the general Kerr-NUT-(A)dS metrics.  }
\end{abstract}

\pacs{04.50.Gh, 04.50.-h, 04.70.-s, 04.20.Jb}
\preprint{Alberta-Thy-02-08}

\maketitle

\section{Introduction}
\subsection{Symmetries}

In modern theoretical physics one can hardly overestimate the  role
of symmetries. They comprise the most fundamental laws of nature, 
they allow us to classify solutions, in their presence  complicated
physical problems become tractable.  The value of symmetries is
especially high in nonlinear theories, such as general relativity. 

In curved spacetime continuous symmetries (isometries)  are generated
by Killing vector fields. Such symmetries have clear geometrical
meaning. Let us assume that in a given manifold we have a
$1$-parameter family of  diffeomorphisms generated by a vector field
$\bss{\xi}$. Such a vector field determines the dragging  of
tensors by the diffeomorphism transformation. If a tensor field $\bss{T}$ is
invariant with respect to this dragging, that is its Lie derivative
along $\bss{\xi}$ vanishes, ${\cal L}_{\xi}\bss{T}=0$, we have a
symmetry. A  vector field which generates transformations
preserving the metric is called a {\em Killing vector field}, and the
corresponding diffeomorphism---an {\em isometry}.  For each of the Killing
vector fields there exists a conserved quantity. For example, for a
particle geodesic motion this conserved quantity is a projection of
the particle momentum on the Killing vector. 

Besides isometries the spacetime may also possess {\em
hidden symmetries}, generated by either symmetric or antisymmetric tensor
fields.  Such symmetries are not directly related to the metric invariance
under diffeomorphism transformations. They
represent the genuine symmetries of the phase space rather than the
configuration space. For example, symmetric Killing tensors  
give rise to conserved quantities of higher order in particle
momenta, and underline the separability of scalar field
equations.  Less known but even more fundamental are antisymmetric
Killing--Yano tensors which are related to the separability of field
equations with spin,  the existence of quantum symmetry operators, and
the presence of  conserved charges.

\subsection{Miraculous properties of the Kerr geometry}

To illustrate the role of hidden symmetries in general relativity let
us  recapitulate the ``miraculous'' properties \cite{Chandra} of the
Kerr geometry. This astrophysically important metric was obtained in
1963 by Kerr \cite{Kerr} as a special solution which can be presented
in the Kerr--Schild form \cite{KerrSchild}
\be\n{KS}
g_{ab}=\eta_{ab}+2H l_{a}l_{b}\, ,
\ee
where $\tens{\eta}$ is a flat metric and $\tens{l}$ is a null vector,
in both metrics $\tens{g}$ and $\tens{\eta}$.\footnote{If the ansatz \eq{KS} is inserted into
the Einstein equations, one effectively reduces the problem to a
linear one (see, e.g., \cite{Gur}). This gives a powerful tool for the study
of special solutions of the Einstein equations. This method works in higher 
dimensions as well. For example, 
the Kerr--Schild ansatz was used by  Myers and Perry
to obtain their higher-dimensional black hole solutions \cite{MP}.}
The Kerr solution is
stationary and axially symmetric, and it belongs to the metrics of
the  special algebraic type {\bf D} of Petrov's classification
\cite{Petrov}.

Although the Killing vector fields $\pa_t$ and $\pa_{\phi}$ are not
enough to provide a sufficient number of integrals of
motion\footnote{For example, for a particle motion these isometries
generate the conserved energy and azimuthal component of the angular
momentum, which together with the conservation of $p^2$ gives only 3
integrals of motion. For separability of the Hamilton--Jacobi
equation in the Kerr spacetime the fourth integral of motion is
required.} in 1968 Carter \cite{Car:68a, Car:68c}
demonstrated that both---the Hamilton--Jacobi and  scalar field
equations---can be separated. This proved, apart from other things,
that there  exists an additional integral of motion, `mysterious'
Carter's constant, which makes the particle geodesic motion
completely integrable. In 1970, Walker and Penrose \cite{WP} pointed
out that Carter's constant is quadratic in particle momenta and its
existence is directly connected with the symmetric Killing tensor
\cite{Stackel}
\be\label{KT}
K_{ab}=K_{(ab)}\,,\quad K_{(ab;c)}=0\,.
\ee
During the several following years it was discovered that it is not
only the Klein--Gordon equation which allows the separation of
variables in the Kerr geometry. In 1972, Teukolsky decoupled the
equations for electromagnetic and gravitational perturbations, and
separated variables in the resulting master equations \cite{Teuk_a}.
One year later the massless neutrino equation by Teukolsky and Unruh
\cite{Teuk_b, Unruh}, and in 1976 the massive Dirac equation by
Chandrasekhar and Page \cite{Chandrasekhar,Page76} were separated.

Meanwhile a new breakthrough was achieved in the field of hidden
symmetries when in 1973 Penrose and Floyd \cite{Penrose} discovered
that the Killing tensor for the Kerr metric can be written in the
form 
\begin{equation}\label{square}
K_{ab}=f_{ac}f^{\
c}_{b}\,,
\end{equation}
where the antisymmetric tensor
$\tens{f}$ is the Killing--Yano (KY) tensor
\cite{Yano} 
\begin{equation}\label{KY4}
f_{ab}=f_{[a b]}\,,\quad f_{a (b;c)}=0\,.
\end{equation}
{A Killing--Yano tensor is in many aspects more fundamental than a
Killing tensor. Namely, its `square' is always Killing tensor, but
in a general case, the opposite is not true \cite{Ferrando}.}

Many of the remarkable properties of the Kerr spacetime are consequences
of the existence of  the Killing--Yano
tensor. In particular, in 1974 Collinson
demonstrated that the integrability conditions for a non-degenerate
Killing--Yano tensor imply that the spacetime is necessary of the
Petrov type {\bf D} \cite{Coll74}.\footnote{All the vacuum type {\bf D} solutions were
obtained by Kinnersley \cite{Kinn}. Demianski and Francaviglia showed
that in the absence  of acceleration these solutions admit
Killing and Killing--Yano tensors \cite{Dem}.} In 1975, Hughston and
Sommers showed that in the Kerr geometry the Killing--Yano
tensor $\tens{f}$ generates both of its isometries 
\cite{kv}. Namely, the Killing vectors $\bss{\xi}$ and $\bss{\eta}$, generating
the time translation and the rotation, can be written as follows:
\begin{equation}\label{kvv}
\xi^a=\frac{1}{3}\,(*f)^{b a}_{\ \ \,;b}=(\partial_t)^{a}\,,\quad
\eta^{a}=-K^{a}_{\ b}\xi^{b}=(\partial_\phi)^{b}\,.
\end{equation}  
This in fact means that all the symmetries necessary for complete
integrability of geodesic motion are `derivable' from the existence
of this Killing--Yano tensor.

In 1977, Carter demonstrated
\cite{Car:77} that given an isometry $\tens{\xi}$ and/or a Killing tensor
$\tens{K}$ one can construct the operators
\begin{equation}\label{opK}
\hat \xi = i\xi^a \nabla_a\,,\quad
\hat K=\nabla_a K^{a b}\nabla_b\,,
\end{equation}
which commute with the scalar Laplacian\footnote{In fact, the operator
$\hat K$ defined by \eqref{opK} commutes with $\Box$ provided that the background metric satisfies the vacuum Einstein  or source-free Einstein--Maxwell equations. In more general spacetimes, however, a quantum anomaly proportional to a contraction of  $\tens{K}$ with the
Ricci tensor may appear.
Such anomaly is not present if 
an additional condition \eqref{square} is satisfied  \cite{cari}.
} 
\begin{equation}
\left[\Box\,, \hat\xi \right]=0=\left[\Box\,, \hat K\right]\,,\quad
\Box=\nabla_a g^{a b}\nabla_b\,.
\end{equation}
Under additional conditions, satisfied by the Kerr geometry, these operators 
commute also between themselves and provide therefore good quantum numbers
for scalar fields. In 1979 Carter and McLenaghan found that an operator
\begin{equation}
\hat f = i\gamma_5\gamma^a\Bigl(f_{a}^{\ b}\nabla_b
-\frac{1}{6}\,\gamma^b\gamma^c f_{a b;c}\Bigr)
\end{equation}
commutes with the Dirac operator $\gamma^a\nabla_a$ \cite{CaMc}. This
gives a new quantum number for the spinor wavefunction and explains 
why  separation of the Dirac equation can be achieved.
Similar symmetry operators for other equations with spin, including
electromagnetic and gravitational perturbations, were constructed
later \cite{spin}. 

In 1983, Marck solved equations for the parallel transport of an
orthonormal frame along geodesics in the Kerr spacetime and used this
result for the study of tidal forces \cite{Marck}. For this construction
he used a simple fact that the vector 
\begin{equation}\label{Lc}
L_a=f_{ab}p^b\,,\quad L_a p^a=0\,,
\end{equation}
is parallel-propagated along a geodesic $\tens{p}$.  

In 1987, Carter \cite{Car:87} pointed out that 
the Killing--Yano tensor itself is derivable from a
$1$-form $\tens{b}\,,$ 
\begin{equation}\label{C87}
\tens{f}=*\tens{d b}\,.
\end{equation} 
Such a form $\tens{b}$ is usually  called   a KY {\em potential}.
It satisfies the Maxwell equations and can be interpreted as a
4-potential of an electromagnetic field with the source current 
proportional to the {\em primary} Killing vector field $\pa_t\,,$ cf. eq.
\eqref{kvv}. In the paper  \cite{f1},  the equations for an
equilibrium configuration of a cosmic  string near the Kerr black
hole were separated. In 1993, Gibbons {\em et al.} demonstrated  that
due to the presence of Killing--Yano tensor the classical spinning
particles in this background possess enhanced worldline supersymmetry
\cite{Gib:93}. Conserved quantities in the Kerr geometry generated by
$\tens{f}$ were discussed in 2006 by Jezierski and \L ukasik
\cite{Jez:06}. 

To conclude this section we  mention that many of the above
statements and results, which we have formulated for the Kerr
geometry, are in fact more general. Their validity can be extended to
more general spacetimes,  or even to an arbitrary number of spacetime
dimensions. For example, the whole Carter's class of solutions
\cite{CDP} admits a KY tensor and possesses many of the discussed
properties.\footnote{A general form of a line element in four dimensions 
admitting a Killing--Yano tensor was  obtained  by Dietz and R\"udiger
\cite{Dietz}.} General results on Killing--Yano tensors  and
algebraic properties were gathered by Hall \cite{Hall}.  
A  relationship among  the existence of Killing tensors,
Killing--Yano tensors, and separability  structures for the
Hamilton--Jacobi equation in arbitrary number of spacetime
dimensions  was discussed in \cite{BF, Dem}.

\subsection{Higher-dimensional black holes}

Recently, a lot of interest focuses on higher-dimensional ($D>4$)
black hole spacetimes. In the widely discussed models with large
extra dimensions it is assumed that one or more additional spatial
dimensions are present. In such models one expects mini black hole
production in the high energy particle collisions \cite{miniBH}. Mini
black holes can serve as a probe of the extra dimensions. At the same
time their interaction with the brane, representing our physical
world, can give the  information about the brane properties. If a
black hole is much smaller that the size of extra dimension and the
brane tension is neglected, its metric can be approximated by an
asymptotically flat or (A)dS solution of the higher-dimensional Einstein
equations.

Study of higher-dimensional black hole solutions has a long history. 
In 1963, Tangherlini \cite{Tang} obtained a higher-dimensional
generalization of the  Schwarzschild metric \cite{Schwarzschild}. 
The charged version of the Tangherlini metric was found  in 1986 by
Myers and Perry \cite{MP}.  In the same paper a general vacuum
rotating black hole in higher dimensions was obtained. This solution,
often called the Myers--Perry (MP) metric,
generalizes the four-dimensional Kerr solution. Main new feature of the MP metrics
in $D$ dimensions  is that instead of 1 rotation parameter, they have
$[(D-1)/2]$ rotation parameters, corresponding to $[(D-1)/2]$
independent $2$-planes of rotation.

Later, in 1998,  Hawking, Hunter, and Taylor-Robinson \cite{HHT}
found a $5$D generalization of the 4D rotating black hole in
asymptotically (anti) de Sitter space (Kerr-(A)dS metric 
\cite{Car:68c}).  In 2004 Gibbons, L\"u, Page, and Pope
\cite{GLPP1,GLPP2} discovered the general Kerr-de Sitter metrics in 
arbitrary number of  dimensions.    After several
attempts to include  NUT \cite{NUT} parameters \cite{CGLP, CLP:B},
in  2006 Chen, L\"u, and Pope \cite{pope}  found a general
Kerr-NUT-(A)dS solution of the Einstein equations for all $D$. 

These metrics were obtained in special coordinates which are
the natural higher-dimensional generalization of 
the Carter's $4$D {\em canonical coordinates} \cite{CDP}. So far 
this metric remains the most
general  black hole type solution of the Einstein 
equations with the cosmological constant (with horizon of
the spherical topology) which is  known
analytically.\footnote{ Besides the brane-world scenarios, these black
holes find their applications in the ADS/CFT correspondence. In the
BPS limit odd dimensional metrics lead to the Sasaki--Einstein
metrics \cite{pope, Sasaki} whereas even dimensional metrics lead to the
Calabi--Yau spaces \cite{Calabi}.
There have been also several attempts to generalize these solutions.
For example, to find a similar solution of the Einstein--Maxwell equations
either in an analytical form \cite{Aliev} or numerically \cite{Kunz}. 
See also \cite{acceleration}.} 
For a recent extended review on higher-dimensional black holes see 
\cite{Em:2008}.

In connection with these black holes a natural question arises: To
what extent are  the remarkable properties of four-dimensional black
holes carried by their higher-dimensional analogues? And in
particular, do these spacetimes possess hidden symmetries?

The hidden symmetries of higher-dimensional rotating black holes were
first discovered  for the $5$D Myers--Perry metrics \cite{FSa,FSb}. It
was demonstrated that both, the Hamilton--Jacobi and  scalar field
equations,  allow the separation of variables and the corresponding
Killing tensor was obtained. 
This, for example, allows one to
obtain a cross-section for the capture of particles and light by 5D black
holes \cite{GoFr}.
 Later it was shown that 5D
results  can be extended to arbitrary number of dimensions,
provided that rotation parameters of the MP metric can be
divided into two classes, and within each of the classes these
parameters are equal one to another. Similar results were found in
the presence of the cosmological constant and NUT parameters
\cite{CGLP, separation}. It was also demonstrated that a stationary
string configuration in the $5$D Myers--Perry spacetime is  
completely integrable  \cite{FrSte}.

\subsection{Recent developments}

Recently, a new breakthrough in the study of 
higher-dimensional rotating black holes was achieved.  It turned out
that the properties of even the most general known higher-dimensional
Kerr-NUT-(A)dS black holes \cite{pope} are, in many aspects, similar
to the properties of their $4$-dimensional `cousins'.  These follow from the existence
of a special closed conformal Killing--Yano (CKY) tensor, 
which is called {\em principal}. 

The principal CKY tensor was first discovered for 
the Myers--Perry metrics \cite{FK}, and soon after that for the completely general Kerr-NUT-(A)dS spacetimes \cite{KF}.
Starting with this tensor, one can generate the
whole {\em tower} of Killing--Yano and Killing tensors \cite{KKPF}
which are responsible for complete integrability of geodesic
motion in these spacetimes \cite{PKVK, KKPV}. Such integrability was
independently proved by separating the Hamilton--Jacobi and
Klein--Gordon equations \cite{FKK}. These results are very
promising and might suggest that also the equations with spin can
be decoupled and separated. The
separation of the Dirac equation was already demonstrated \cite{oy}. 
Also worth mentioning is a recent work on separability of vector
and tensor fields in  $D=5$ Myers--Perry  spacetimes with equal
angular momenta \cite{Murata}, and proved  complete integrability of
stationary string configurations in general Kerr-NUT-(A)dS spacetimes
\cite{KF_string}. For a brief review of these results see \cite{Frolov}.

In this paper, we discuss the hidden symmetries of rotating  black
holes.  The following section contains basic definitions. In
Section~III, starting from a 4D flat space and choosing a special
principal CKY tensor, we `derive' the $4$D Kerr-NUT-(A)dS spacetime
and its hidden symmetries. In Section~IV we prove the central
theorem concerning the properties of closed CKY tensors and introduce
the tower of Killing and Killing--Yano tensors in higher-dimensional spacetimes.
We also construct
the canonical basis associated with the principal CKY tensor.  In
Section~V we apply these results to higher-dimensional 
Kerr-NUT-(A)dS solutions. Section~VI contains discussion of the
separability problem. Possible future developments are discussed in
Section~VII.

\section{Killing--Yano and Killing tensors}

\subsection{Definitions}

Let us consider a $D$-dimensional spacetime with a metric $\bss{g}$.
In order to simultaneously cover both cases of odd and even dimensions 
we write $D=2n+\varepsilon$,
where $\ve=0$ ($\ve=1$) for the even (odd) dimensional case. 
A spacetime  possesses an isometry generated by the
Killing vector field $\tens{\xi}$ if this vector obeys the {\em
Killing equation} 
\be 
\xi_{(a;b)}=0\, . 
\ee 
For a geodesic motion of a particle in such curved spacetime the quantity
$p^{a}\xi_{a}$, where $\tens{p}$ is the momentum of the particle, remains constant
along the particle's trajectory.
Similarly, for a null geodesic,
$p^{a}\xi_{a}$ is conserved provided $\tens{\xi}$ is a {\em conformal
Killing vector} obeying the equation
\be
\xi_{(a;b)}=\tilde{\xi}g_{ab}\hh \tilde{\xi}=D^{-1}\xi^{b}_{\ \ ;b}\, .
\ee

There exist two natural (symmetric and antisymmetric) generalizations
of a (conformal) Killing vector.

A symmetric (rank-$p$) {\em conformal Killing tensor} \cite{WP}
$\tens{K}$ obeys the equations
\be\n{CK}
K_{a_1a_2 \ldots a_p}=
K_{(a_1a_2 \ldots a_p)}\,,\quad 
K_{(a_1a_2 \ldots a_p;b)}=g_{b (a_1}\tilde{K}_{a_2 \ldots
a_p)}
\, .
\ee
As in the case of a conformal Killing vector, the tensor
$\tilde{\bss{K}}$ is  determined by tracing both sides of equation
\eq{CK}. If $\tilde{\bss{K}}$ vanishes, the tensor $\bss{K}$ is called
a {\em Killing tensor} \cite{Stackel}. In a presence of the Killing
tensor $\bss{K}$ the conserved quantity for a geodesic motion is 
\begin{equation}
K=K_{a_1a_2 \ldots a_p}p^{a_1}p^{a_2}\ldots p^{a_p}\,.
\end{equation}
For null geodesics this quantity is conserved not only for a Killing
tensor, but also for a conformal Killing tensor.

A {\em conformal Killing--Yano} (CKY) tensor \cite{Tachibana, cari} 
$\tens{h}$ is an {\em antisymmetric} tensor
\be\label{PTL}
h_{a_1a_2 \ldots a_p}=\ h_{[a_1a_2 \ldots a_p]}\,
\ee
which obeys 
\be\label{CKY}
\nabla_{(a_1}h_{a_2)a_3 \ldots a_{p+1}}=\ 
g_{a_1a_2}\tilde{h}_{a_3 \ldots a_{p+1}}-
(p-1)g_{[a_3(a_1}\tilde{h}_{a_2) \ldots a_{p+1}]}\, .
\end{equation}
By tracing both sides of this equation one obtains the following
expression for $\tilde{\bss{h}}$
\be
\tilde{h}_{a_2 a_3 \ldots a_{p}}={1\over D-p+1}\nabla^{a_1}h_{a_1a_2
\ldots a_p}\, .
\ee
In the case when $\tilde{\bss{h}}=0$ one has a {\em  Killing--Yano}
(KY) tensor \cite{Yano}. For the KY tensor $\bss{h}$ the quantity
\begin{equation}\label{Lprop}
L_{a_1a_2 \ldots a_{(p-1)}}=h_{a_1a_2 \ldots a_p}p^{a_p}\,, 
\end{equation}
is parallel-propagated along the geodesic $\tens{p}$.

Let us mention two additional important properties:
having a KY tensor $\tens{h}$
the quantity
\begin{equation}\label{CKT} 
K_{ab}=\frac{c_p}{(p-1)!}\,h_{a a_2 \ldots
a_p}h_{b}^{\ \,a_2 \ldots
a_p}
\end{equation}
is an associated Killing tensor. Here $c_p$ is an arbitrary constant,
which is often taken to be one. For a different convenient choice
see Section~IV.  For a CKY tensor $\bss{h}$ of rank-$2$ the vector
\be\n{prime}
\xi^{(0)a}={1\over D-1}\nabla_b h^{ab}
\ee
obeys the following equation \cite{jez}
\be\label{kvjez}
\xi^{(0)}_{\ (a;b)}=-{1\over D-2}R_{c (a}h_{b)}^{\,\,\,\,c}\,
.
\ee
Thus, in an Einstein space, that is when
$R_{ab}=\Lambda g_{ab}$, $\bss{\xi}^{(0)}$ is the Killing vector.

\subsection{Killing--Yano equations in terms of differential forms}

The CKY tensors are forms and operations with them are greatly
simplified if one uses the `language' of differential forms. 

Let us remind some of the relations we shall use later.
If $\tens{\alpha}_p$ and $\tens{\beta}_q$ are $p$- and $q$-forms,
respectively, the external derivative $\tens{d}$ of their exterior product
$\wedge$ obeys a relation
\be\n{dd}
\tens{d}(\bss{\alpha}_p\wedge \bss{\beta}_q)=
\tens{d}\bss{\alpha}_p\wedge \bss{\beta}_q+
(-1)^p\bss{\alpha}_p\wedge \tens{d}\bss{\beta}_q\, .
\ee
A
Hodge dual $*\bss{\alpha}_p$ of the $p$-form $\bss{\alpha}_p$ is
$(D-p)$-form defined as 
\be
(* {\alpha}_p)_{a_1 \ldots a_{D-p}}= {1\over p!}\, \alpha^{b_1\ldots
b_p}e_{b_1\ldots b_p a_1 \ldots a_{D-p}}\, ,
\ee
where $e_{a_1 \ldots a_D}$ is a totally antisymmetric tensor.
The {\em co-derivative}
$\tens{\delta} $ is defined as follows:
\be
\tens{\delta} \bss{\alpha}_p=(-1)^p\epsilon_p *\tens{d}*\bss{\alpha}_p\,,\quad
\epsilon_p=(-1)^{p(D-p)}{\mbox{det}(g)\over |\mbox{det}(g)|}\, .
\ee
One also has $* *\bss{\alpha}_p=\epsilon_p\bss{\alpha}_p$.

If $\{ \bss{e}_a\}$ is a basis of vectors, then dual basis of 1-forms
$\bss{\omega}^a$ is defined by the relations
$\bss{\omega}^a(\bss{e}_b)=\delta^a_b$. We denote
$\eta_{ab}=\tens{g}(\bss{e}_a,\bss{e}_b)$ and by $\eta^{ab}$ the inverse
matrix. The operations with the indices enumerating the basic
vectors and forms are performed by using these matrices. In
particular, $\bss{e}^a=\eta^{ab}\bss{e}_b$, and so on. We denote a
covariant derivative along the vector $\bss{e}_a$ by
${\nabla}_a={\nabla}_{{e}_a}$.
 One has
\be\label{2.15}
\tens{d}=\bss{\omega}^a\wedge {\nabla}_a\,,\quad
\tens{\delta}=-\bss{e}^a\hook {\nabla}_a\, .
\ee
In tensor notations the `hook' operator (inner derivative) 
along a vector $\tens{X}$, applied to a $p$-form
$\bss{\alpha}_p$, corresponds to a contraction 
\be
({X}\hook {\alpha}_p)_{a_2 \ldots a_p}= X^{a_1} (\alpha_p)_{a_1 a_2
\ldots a_p}\, .
\ee
It satisfies the properties
\begin{equation}\label{hook2}
\tens{e}^a \hook (\tens{\alpha}_p\wedge \tens{\beta}_q)=
(\tens{e}^a \hook \tens{\alpha}_p)\wedge \tens{\beta}_q +(-1)^p
\tens{\alpha}_p\wedge (\tens{e}^a \hook \tens{\beta}_q)\,,
\end{equation}
\be\label{hook1}
\bss{e}^a\hook \bss{\omega}_a=D\,,\quad 
\bss{\omega}_a\wedge (\bss{e}^a\hook \bss{\alpha}_{p})=p\,\bss{\alpha}_{p}\,.
\ee

For a given vector $\bss{X}$ one  defines $\bss{X}^{\flat}$   as a
corresponding 1-form with the components
$(X^{\flat})_{a}=g_{ab}X^{b}$. 
In particular, one has $\eta^{ab}(\bss{e}_b)^{\flat}=\bss{\omega}^a$.
We refer to \cite{sten,kress} where  these and many other useful
relations can be found.

Definition \eq{CKY} of the CKY tensor $\bss{h}$ (which is a
$p$-form) is equivalent to the following equation \cite{kress}:
\be\n{CKYf}
{\nabla}_{X} \bss{h}={1\over p+1} \bss{X}\hook \tens{d}\bss{h}
-{1\over D-p+1}\bss{X}^{\flat}\wedge
\tens{\delta} \bss{h}\, .
\ee 
That is, a CKY tensor is a form for which the covariant derivative 
splits into the exterior and divergence parts.\footnote{
For a general form an additional term, the harmonic part, is present.
It is the lack of this term what makes CKY tensors `special'.}
Using the relation
\be\label{definition}
\bss{X}\hook *\bss{\omega}=*(\bss{\omega}\wedge \bss{X}^{\flat})\,, 
\ee
it is easy to show that under the Hodge duality the exterior part 
transforms into the divergence part and vice versa. In particular,
\eq{CKYf} implies
\be
{\nabla}_{X}(* \bss{h})={1\over p_*+1} \bss{X}\hook \tens{d}(*\bss{h})
-{1\over D-p_*+1}\bss{X}^{\flat}\wedge
\tens{\delta}(*\bss{h})\,,\quad p_*=D-p\, .
\ee
The Hodge dual $*\bss{h}$ of a CKY tensor $\bss{h}$ is again
a CKY tensor. 

Two special subclasses of CKY tensors are of particular interest:
(a) Killing--Yano tensors with zero divergence part $\tens{\delta} \tens{h}=0$
and (b) closed CKY tensors with vanishing exterior part $\tens{dh}=0$.
Under the Hodge duality these subclasses transform into each other.

For a closed CKY tensor there exists locally a (KY) potential $\tens{b}$,
which is a $(p-1)$-form, such that 
\begin{equation}\label{db}
\tens{h}=\tens{db}\,.
\end{equation}
The Hodge dual of such a tensor $\bss{h}$,
\begin{equation}
\tens{f}=*\tens{h}=*\tens{db}\, ,
\end{equation}
is a Killing--Yano tensor, cf. eq. \eqref{C87}.

\section{4D Kerr-NUT-(A)dS spacetime and its hidden symmetries}

Before we proceed to higher-dimensional rotating black holes  and
their hidden symmetries  it is instructive to illustrate the basic
ideas on the well known $4$D case. As we shall see later, a key
object of the theory in higher dimensions is  a {\em principal CKY
tensor}.  We start discussing this object and its properties in $4$D
flat spacetime and demonstrate how it generates other objects
(Killing--Yano and Killing tensors) responsible for the hidden
symmetries.  We also show how this principal CKY tensor allows one
easily to `generate' the $4$D Kerr-NUT-(A)dS metric starting from the
flat one---written in the {\em canonical coordinates} determined by
this tensor.\footnote{For an alternative `derivation' of the Kerr-NUT-(A)dS 
spacetime see, e.g., \cite{Dadhich} or \cite{Misha}.} We also demonstrate the separation
of variables in the $4$D Kerr-NUT-(A)dS spacetime in the canonical coordinates.
It should
be emphasized, that this section plays the role of an introduction
for newcomers to the field which should illuminate the main ideas of more
complicated higher-dimensional theory.

\subsection{Principal conformal Killing--Yano tensor}

Consider a four-dimensional flat spacetime with the metric
\begin{equation}
dS^2=\eta_{ab}dX^adX^b=-dT^2+dX^2+dY^2+dZ^2\,.
\end{equation}
The principal CKY tensor $\tens{h}$ is the rank-$2$ closed CKY
tensor. Therefore, there exists a $1$-form potential $\tens{b}$ so
that $\tens{h}=\tens{db}$. Let us consider the following ansatz:
\be\label{bb}
\bss{b}={1\over 2}\,\bigl[-R^2\tens{d}T+a(Y\tens{d}X-X\tens{d}Y)\bigr]\,,\quad 
R^2=X^2+Y^2+Z^2\, .
\ee
Our choice of this special form for the
potential $\tens{b}$ will become clear later, when it will be shown that this is
a flat spacetime limit of the potential for the principal CKY
tensor in the Kerr-NUT-(A)dS spacetime. For a moment we just mention
that the form \eq{bb} of the potential $\bss{b}$ singles out time coordinate $T$, 
a two-dimensional $(X,Y)$ plane in space, and contains an arbitrary constant $a$.

It can be easily shown that 
\be\label{hh}
\bss{h}=\bss{db}=\tens{d}T\wedge(X \tens{d}X+Y \tens{d}Y+Z\tens{d}Z)+a \tens{d}Y
\wedge \tens{d}X\, 
\ee
is a closed conformal Killing--Yano tensor. 
It means that its dual 2-form $\bss{f}=*\bss{h}$ is the Killing--Yano tensor.\footnote{
In $D$ dimensions the maximum number of (linear independent) Killing--Yano tensors of a given rank-$p$ is  
\be
N_p=\left(\!\!   
\begin{array}{c}
D \\
p\end{array}
\!\!\right)\,+ 
\left(\!\!   
\begin{array}{c}
D \\
p+1\end{array}
\!\!\right)\,=\frac{(D+1)!}{(D-p)!(p+1)!}.
\ee
This reflects the fact that, similar to Killing vectors,  
Killing--Yano tensors are completely determined by the values of
their components and the values of their (completely antisymmetric) 
first derivatives at a given point. Flat space has the maximum number of independent 
Killing--Yano tensors of each rank. Any KY tensor there can be written as 
a linear combination of  `translational' KY tensors (which are a simple wedge
product of
translational Killing vectors) and `rotational'  KY tensors (which
are a wedge product of translations with a spacetime rotation,
completely antisymmetrized) \cite{KaTr}. In particular case of $D=4$
we have 10 rank-$2$ KY tensors (6 translational and 4 rotational).
}
\begin{equation}\label{ff}
\tens{f}=X \tens{d}Z\wedge \tens{d}Y+Z \tens{d}Y\wedge \tens{d}X+
Y \tens{d}X\wedge \tens{d}Z+a \tens{d}Z\wedge \tens{d}T\,.
\end{equation}

Let us put, for a moment, $a=0$. Then the KY tensor $f_{ab}$ has only
spatial components $f_{ik}\,,$ and the Killing tensor \eqref{CKT}, associated with it,  reads
\begin{equation}
K_{ij}=R^2\delta_{ij}-X^iX^j=\!\!\sum_{k=X,Y,Z} \xi_{(k)\,i}\xi_{(k)\,j}\,,
\quad \xi_{(k)\,i}=\epsilon_{kji}X^j\,.
\end{equation}
Here ${\xi}_{(k)\,i}$ are the spatial rotational Killing vectors.
Therefore, the Killing tensor $\tens{K}$ can be written as a sum of
products of  Killing vectors, and thus it is {\em reducible}.
Parallel-propagated vector \eqref{Lprop}
\begin{equation}
L_{i}=f_{ik}p^k=\epsilon_{ijk}X^jp^k={\xi}_{(k)\,i}p^k
\end{equation}
has the meaning of the conserved angular momentum.\footnote{ In
general, for a simple spacelike ($f_{ab}f^{ab}>0$) Killing--Yano
tensor  $\tens{f}$, there exists a close analogy between the angular
momentum of classical mechanics and the vector $L^a=f^{ab}p_{b}$
\cite{Dietz}.}  The conserved quantity
\begin{equation}
K(a=0)=\!\!\sum_{k=X,Y,Z} L^2_{k}={\vec L}^2
\end{equation}
is the square of the total angular momentum.

For $a\neq 0$ the conserved quantity
\begin{equation}\label{K}
K={\vec L}^2+2aEL_{Z}+a^2(E^2-p_{Z}^2)\,,
\end{equation}
is also reducible. Here $E=-p_T$ and
$p_{Z}$ are the conserved energy and the momentum
in the $Z$-direction, respectively. 

\subsection{`Derivation' of the 4D Kerr-NUT-(A)dS metric} 

Consider a general case with $a\ne 0$. We first introduce the ellipsoidal
coordinates\footnote{In this step, we associate  constant $a$ with
`rotation' parameter.} 
\begin{equation}\n{eq_3.15}
X=\sqrt{r^2+a^2}\sin\theta\cos\phi,\quad 
Y=\sqrt{r^2+a^2}\sin\theta\sin\phi,\quad 
Z=r\cos\theta\,, 
\end{equation}
and rewrite the metric, the potential, the principal CKY tensor, and the KY tensor as 
\begin{equation}
\begin{split}
dS^2=&\,-dT^2+(r^2+a^2)\sin^2\!\theta d\phi^2+(r^2+a^2\cos^2\theta)
\bigl(\frac{dr^2}{r^2+a^2}+d\theta^2\bigr)\,,\\
\tens{b}=&\,\frac{1}{2}\bigl[-(r^2+a^2\sin^2\!\theta)\tens{d}T-a\sin^2\!\theta(r^2+a^2)\tens{d}\phi\bigr]\,,\\
\tens{h}=&\,-r \tens{d}r\wedge(\tens{d}T+a\sin^2\!\theta \tens{d}\phi)-a\sin\theta \cos\theta \tens{d}\theta\wedge
\bigl[a\tens{d}T+(r^2+a^2)\tens{d}\phi\bigr]\,,\\
\tens{f}=&\,a\cos\theta \tens{d}r\wedge(\tens{d}T+a\sin^2\!\theta \tens{d}\phi)-r\sin\theta \tens{d}\theta\wedge
\bigl[a\tens{d}T+(r^2+a^2)\tens{d}\phi\bigr]\,.
\end{split}
\end{equation}
Second, we introduce the new coordinates  
\begin{equation}\n{eq_3.18}
y=a\cos\theta,\quad t=T+a\phi,\quad \psi=-\phi/a\,,
\end{equation}
in which the metric takes the `algebraic' form 
\begin{equation}\label{CDP}
dS^2=-\frac{\Delta_r (dt+y^2d\psi)^2}{r^2+y^2}+\frac{\Delta_y (dt-r^2d\psi)^2}{r^2+y^2}
+\frac{(r^2+y^2)dr^2}{\Delta_r}+\frac{(r^2+y^2)dy^2}{\Delta_y}\,,
\end{equation}
where 
\begin{equation}\label{ry}
\Delta_r=r^2+a^2,\quad 
\Delta_y=a^2-y^2\,.
\end{equation}
The hidden symmetries are
\begin{equation}\label{form}
\begin{split}
\tens{b}=&\,\frac{1}{2}\bigl[(y^2-r^2-a^2)\tens{d}t-r^2 y^2 \tens{d}\psi\bigr]\,,\\
\tens{h}=&\,y\tens{d}y\wedge(\tens{d}t-r^2\tens{d}\psi)-r\tens{d}r\wedge(\tens{d}t+
y^2\tens{d}\psi)\,,\\
\tens{f}=&\,r\tens{d}y\wedge(\tens{d}t-r^2\tens{d}\psi)+y\tens{d}r\wedge(\tens{d}t+y^2\tens{d}\psi)\,.
\end{split}
\end{equation}
In the potential $\tens{b}$ the term proportional to $a^2$ is constant and may be omitted.
We remind that \eqref{CDP}--\eqref{ry} is just a metric of the flat
space written in special coordinates. 

Let us consider now the metric \eq{CDP} without imposing 
conditions \eq{ry} on functions $\Delta_r$ and
$\Delta_y$, but assuming that they are functions of $r$ and $y$,
respectively. Then substituting this {ansatz} into the Einstein
equations
\be\n{eqq}
R_{ab}=-3\lambda g_{ab}\, ,
\ee
one finds that these equations are satisfied provided the following
relation is valid:
\be
{d^2{\Delta}_r\over dr^2}+ {d^2{\Delta}_y\over dy^2}=12\lambda(r^2+y^2)\, .
\ee
The most general solution of this equation is 
\be\label{functions}
\Delta_r=(r^2+a^2)(1+\lambda r^2)-2Mr\hhh
\Delta_y=(a^2-y^2)(1-\lambda y^2)+2Ny\,.
\ee 
In other words, a simple
replacement of functions 
\eqref{ry} by more general polynomials \eq{functions} generates a
non-trivial solution of the Einstein equations from a flat one. This
solution is  the Kerr-NUT-(A)dS  metric 
written in the {\em canonical form} \cite{CDP}.
It obeys the Einstein equations
with the cosmological constant (see \eq{eqq}). $M$ stands for mass, and
parameters $a$ and $N$ are connected with rotation and NUT
parameter \cite{GP}.  

A remarkable fact is that in canonical coordinates, with arbitrary
$\Delta_r(r)$ and $\Delta_y(y)$, the objects $\bss{b}$, $\tens{h}$, and $\bss{f}$ 
\eq{form} are again  the potential, the principal CKY tensor, and the KY tensor
for the metric \eq{CDP} \cite{KF, KuKr}. In particular,
these relations give a principal CKY tensor, and a
derived from it $2$-form of the KY tensor, for the
Kerr-NUT-(A)dS  spacetime \eqref{CDP}, \eqref{functions}.

Let us emphasize that in the Kerr-NUT-(A)dS spacetime neither
the square of the total angular momentum, ${\vec L}^2$, nor the projection of the
momentum on the $Z$-axis, $p_Z$, which enter \eq{K} have well defined
meaning. However, the quadratic in momentum quantity, $K^{ab}p_ap_b$,
where $K_{ab}=f_{ac}f_b^{\ c}$, is well defined and conserved. In the
absence of the cosmological constant and NUT parameter, that is for
the Kerr black hole, this quantity
can be presented in the form \eq{K} in the asymptotic region, where the spacetime is practically
flat. The angular momentum
and other quantities which enter \eq{K} must be then understood as the
corresponding asymptotically conserved quantities. Since the energy
$E$ and the angular momentum along the axis of symmetry $L_Z$ are conserved exactly in any stationary axisymmetric
spacetime they can be excluded from \eq{K} and the asymptotically
conserved quantity can be written as follows \cite{RBS}:
\be
Q=L_{X}^2+L_{Y}^2-a^2p_Z^2\, .
\ee
For a scattering of particles in the Kerr metric, the presence of an
exact integral of motion connected with the Carter's constant implies
that the quantity $Q$ calculated for the incoming from infinity
particle must be the same as $Q$ calculated at the infinity for the
outgoing particle. An interesting question is the following: suppose that 
such a conservation law is established for any scattering of particles by a
localized object, can one conclude that the metric of this object
possesses a hidden symmetry?

\subsection{Symmetric form of the metric}

Let us perform the `Wick' rotation in radial coordinate $r$. This  
transforms the metric \eqref{CDP} and its hidden symmetries into 
a {\em symmetric form} \cite{pope}.
After transformation
\begin{equation}\n{wr}
r=ix\,,\quad M=iN_x\,,\quad N=N_y\,,
\end{equation}
the metric and the KY objects take the form
\begin{equation}\label{wick}
ds^2=\frac{\Delta_x (dt+y^2d\psi)^2}{x^2-y^2}+\frac{\Delta_y (dt+x^2d\psi)^2}{y^2-x^2}
+\frac{(x^2-y^2)dx^2}{\Delta_x}+\frac{(y^2-x^2)dy^2}{\Delta_y}\,,
\end{equation}
\begin{equation}\label{deltawick}
\Delta_x=(a^2-x^2)(1-\lambda x^2)+2N_xx\,,\ 
\Delta_y=(a^2-y^2)(1-\lambda y^2)+2N_yy\,,
\end{equation}
\begin{equation}\n{eq_3.29}
\tens{b}=\frac{1}{2}\bigl[(x^2+y^2)\tens{d}t+x^2y^2\tens{d}\psi\,\bigr]\,,
\end{equation}
\begin{equation}\label{hwick}
\tens{h}=y\tens{d}y\wedge(\tens{d}t+x^2\tens{d}\psi)+x\tens{d}x\wedge(\tens{d}t+y^2\tens{d}\psi)\,.
\end{equation}
\begin{equation}\label{fwick}
\tens{f}=x\tens{d}y\wedge(\tens{d}t+x^2\tens{d}\psi)+y\tens{d}x\wedge(\tens{d}t+y^2\tens{d}\psi)\,.
\end{equation}
This form of the Kerr-NUT-(A)dS spacetime and of the potential (the principal CKY
tensor) allows a natural
generalization to higher dimensions \cite{pope, KF}.\footnote{
It is obvious from the derivation that this symmetric form of the metric and of
its hidden symmetries is an analytical continuation of the real physical
quantities \eqref{CDP}, \eqref{form}, \eqref{functions}. The signature of the metric for 
this continuation depends on the domain of coordinates $x$ and $y$ and the signature of $\Delta_x$ and $\Delta_y$. For example, for $x>y$ and
$\Delta_x>0$, $\Delta_y<0$ it is of the Euclidean signature. The
transition to the physical space is given by \eqref{wr}. 
In higher dimensions it is very convenient to work with
a generalization of this symmetric form. }

\subsection{Principal conformal Killing--Yano tensor and canonical
coordinates}

We demonstrate now that the coordinates $(t,x,y,\psi)$ used in 
\eq{wick}--\eq{fwick} have a deep invariant meaning.  Let us define 
\be
H^a_{\ b}=h^{a c}h_{b c}\hh
\Delta^a_{\  b}=H^a_{\ b}-H \delta^a_{\  b}\, ,
\ee
then one has
\be
\Delta^a_{\ b} = \left( 
{\begin{array}{cccc}
-R^2 - H  & a \,Y &  - a \,X & 0 \\
 - a \,Y & a^2-X^{2} - H  & -Y\,X & -Z\,X \\
a \,X & -Y\,X & a^2-Y^{2} - H  & -Z\,Y \\
0 & -Z\,X & -Z\,Y & -Z^{2} - H 
\end{array}}
 \right) \, .
\ee
The condition $\mbox{det} (\Delta)=0$ which determines the eigenvalues $H$
of the operator $\bss{H}$\footnote{The operator $\tens{H}$ is the conformal Killing tensor. It is related to $\tens{K}$ as 
\be
K_{ab}=H_{ab}-\frac{1}{2}g_{ab}H^c_{\ c}\,,\qquad
H_{ab}=K_{ab}-\frac{1}{D-2}\,g_{ab}K^c_{\ c}\,.
\ee
}  
is equivalent to the following equation:
\be
H^{2}+(R^2-a^2)H -a ^{2}\,Z^{2}=0\, .
\ee
Hence the eigenvalues of $\bss{H}$ are
\be
H_{\pm}={1\over 2}\Bigr[ a^2-R^2\pm\sqrt{(R^2-a^2)^2+4a^2Z^2}\Bigr]\, .
\ee
Simple calculations using \eq{eq_3.15} give
\be
H_+=a^2\cos^2\theta=y^2\,,\quad
H_-=-r^2=x^2\,.
\ee
Thus the coordinates $x$ and $y$ in \eq{wick} are uniquely
determined as the eigenvalues of the operator $\bss{H}$ constructed
from the principal CKY tensor $\bss{h}$. Let us show
now that the same tensor $\bss{h}$ uniquely determines the coordinates
$t$ and $\psi$. The primary Killing vector $\bss{\xi}^{(0)}$,
\eq{prime}, in our case
is 
\be
\bss{\xi}^{(0)}=\pa_T\, .
\ee
Moreover, 
\be
{\xi}^{(\psi)\,a}=-K^{ab}{\xi}^{(0)}_{\ b}=a^2(\partial_{T})^a+
aY(\partial_X)^a-aX(\partial_Y)^a\, \, ,
\ee
is the {\em secondary} Killing vector, cf. eq \eqref{kvv}. 
In coordinates \eq{eq_3.18} one has
\be
\bss{\xi}^{(0)}=\pa_t\hh
\bss{\xi}^{(\psi)}=\pa_{\psi}\, .
\ee
It means that the coordinates $t$ and $\psi$ are the affine parameters
along the primary and secondary Killing vectors $\bss{\xi}^{(0)}$ and $\bss{\xi}^{(\psi)}$, determined by the tensor $\bss{h}$.
It can be checked that the
same is true for the Kerr-NUT-AdS metric \eq{wick}, \eq{deltawick} with the principal  CKY tensor $\tens{h}$ given by \eqref{hwick}. This underlines the exceptional 
role of this tensor. Remarkably, the existence of a similar object in the higher-dimensional Kerr-NUT-(A)dS spacetime generates besides the tower of hidden symmetries also all the isometries of this spacetime, in a way exactly analogous to 
four dimensions (see Section V). It also determines the canonical coordinates for these metrics \cite{KKPF, hoy1, hoy2}.

\subsection{Separation of variables}

The last subject we would like to discuss in this brief review of
properties of the 4D Kerr-NUT-(A)dS metrics is the separation of
variables for the Hamilton--Jacobi and Klein--Gordon equations. 

Let us first consider the Klein--Gordon equation
\be\n{KG}
\Box \Phi-\mu^2\Phi =0\, 
\ee
in a spacetime \eq{wick} with $\Delta_x$ and $\Delta_y$  
arbitrary functions of $x$ and $y$, respectively. 
The separation of variables of equation \eq{KG} in
the canonical coordinates $(\tau,x,y,\psi)$ means that $\Phi$ can be
decomposed into modes
\be
\Phi=e^{i\varepsilon \tau+im\psi} X(x) Y(y)\, .
\ee
Indeed, substituting this expression in the Klein--Gordon equation \eq{KG} one
obtains
\be\label{sep1}
(\Delta_x X')'+V_x X=0\,,\quad
V_x=\kappa+\mu^2 x^2-{ (\varepsilon x^2-m)^2\over \Delta_x}\, ,
\ee
\be\label{sep2}
(\Delta_y Y')'+V_y Y=0\,,\quad
V_y=\kappa+\mu^2 y^2-{ (\varepsilon y^2-m)^2\over \Delta_y}\, .
\ee
Here the prime stands for the derivative of function with respect to 
its single argument. The separation constants $\varepsilon$ and $m$ are connected
with the symmetries generated by the Killing vectors
$\bss{\xi}^{(\tau)}=\pa_{\tau}$ and 
$\bss{\xi}^{(\psi)}=\pa_{\psi}$. An additional separation constant
$\kappa$ is connected with the hidden symmetry generated by the Killing
tensor $\bss{K}$.
It should be emphasized, that in order to use the proved 
separability for concrete calculations in the physical Kerr-NUT-(A)dS
spacetime \eq{CDP}, \eq{functions}, one needs to specify functions $\Delta_x$
and $\Delta_y$ to have the form \eqref{deltawick} and perform the Wick transformation
inverse to \eq{wr}. This transformation `spoils' the symmetry between
the essential coordinates but the separability property remains.
In coordinates $r$ and $y$ in the `physical' sector equations 
\eq{sep1} and \eq{sep2} play different roles. Eq. \eq{sep2} with imposed
regularity conditions serves as an eigenvalue problem which determines the spectrum
of $\kappa$. Eq. \eq{sep1} is a radial equation for propagating modes.

Similarly, the Hamilton--Jacobi equation for geodesic motion
\be
\partial_{\lambda} S+g^{ab}\partial_{a} S\partial_{b} S=0
\ee
in a generalized metric \eq{wick} allows a separation of variables and
$S$ can be written in the form
\be
S=\mu^2\lambda+\varepsilon \tau+m\psi+S_x(x)+S_y(y)\, .
\ee
The functions $S_x$ and $S_y$ obey the equations
\be
({S'}_x)^2={V_x\over \Delta_x}\hhh
({S'}_y)^2={V_y\over \Delta_y}\, .
\ee

Presented `derivation' of 4D Kerr-NUT-(A)dS
metric and its hidden symmetries can be naturally generalized into 
higher dimensions. We shall not do this here, but simply mention that
the corresponding expressions for the flat spacetime metric in the
canonical coordinates and for the potential $\bss{b}$ can be easily
obtained by taking the flat space limit of the formulae 
\eq{bas_e} and \eq{bbbb} (see Section~V).

\section{Towers of Killing and Killing--Yano tensors}

\subsection{Important property of closed CKY tensors}

The following result \cite{KKPF} plays a central role in the
construction of the hidden symmetry objects in higher-dimensional
spacetimes. 

{\bf Theorem.} {\em Let $\bss{h}^{(1)}$ and $\bss{h}^{(2)}$ be two closed CKY
tensors. Then their external product $\bss{h}=\bss{h}^{(1)}\wedge
\bss{h}^{(2)}$ is also a closed CKY tensor.}

We shall prove this Theorem in two steps. The fact that $\tens{h}$ is closed is trivial, cf. eq. \eq{dd}. Let us first show that for a $p$-form $\bss{\alpha}_p$ obeying the equation
\be
\nabla_X\bss{\alpha}_p=\bss{X}^{\flat}\wedge \bss{\gamma}_{p-1}\,,
\ee
one has
\be\n{eq0}
\tens{\gamma}_{p-1}=-{1\over D-p+1}\,\tens{\delta}\bss{\alpha}_p\, .
\ee 
Indeed, we find
\ba\n{eq2}
-\tens{\delta}\bss{\alpha}_p&=&\bss{e}^a \hook \nabla_a \bss{\alpha}_p=
\bss{e}^a \hook (\bss{\omega}_a \wedge \bss{\gamma}_{p-1})\nonumber\\
&=&(\bss{e}^a\hook \bss{\omega}_a)\bss{\gamma}_{p-1}
-\bss{\omega}_a\wedge(\bss{e}^a\hook
\bss{\gamma}_{p-1})=(D-p+1)\bss{\gamma}_{p-1}\, .
\ea
Here we have used eq. \eqref{2.15},
and relations 
\eqref{hook2}, \eqref{hook1}.

The second step in the proof of the Theorem is to show that if
$\bss{\alpha}_p$ and $\bss{\beta}_q$ are two closed CKY tensors then
\be
\nabla_X(\bss{\alpha}_p\wedge
\bss{\beta}_q)=\bss{X}^{\flat}\wedge\bss{\gamma}_{p+q-1}\, .
\ee
Really, one has
\be\n{eq3}
\begin{split}
\nabla_X(\bss{\alpha}_p\wedge
\bss{\beta}_q)=&\ 
\nabla_X\bss{\alpha}_p\wedge
\bss{\beta}_q+
\bss{\alpha}_p\wedge
\nabla_X\bss{\beta}_q\\
=&\, -{1\over D-p+1}(\bss{X}^{\flat}\!\wedge \tens{\delta} \bss{\alpha}_p)\wedge
\bss{\beta}_q
-{1\over D-q+1}
\bss{\alpha}_p\wedge (\bss{X}^{\flat}\!\wedge \tens{\delta}
\bss{\beta}_q)\\
=&\ \bss{X}^{\flat}\wedge \bss{\gamma}_{p+q-1}\,,
\end{split}
\ee
where
\be
\bss{\gamma}_{p+q-1}=-{1\over D-p+1} \tens{\delta} \bss{\alpha}_p\wedge
\bss{\beta}_q
-{(-1)^p\over D-q+1}
\bss{\alpha}_p\wedge \tens{\delta}
\bss{\beta}_q\, .
\ee
Combining \eq{eq3} with \eq{eq0} we arrive at the statement of the theorem.

\subsection{Principal CKY tensor and towers of hidden symmetries}
Let us
consider now a special case which is important for applications.
Namely, we assume that a spacetime allows a 2-form $\bss{h}$ which is a
{\em closed CKY tensor}, $\bss{h}=\bss{db}$.  We also assume that
$\tens{h}$ is {\em non-degenerate}, that is its (matrix) rank is $2n$. We
call this object a {\em principal CKY tensor}  \cite{KKPF}. 

According to the
Theorem of the previous section, the principal CKY tensor
generates a set (tower) of new closed CKY tensors
\be\label{hj}
\bss{h}^{(j)}=\bss{h}^{\wedge j}=\underbrace{\bss{h}\wedge \ldots \wedge
\bss{h}}_{\mbox{\tiny{total of $j$ factors}}}\, .
\ee 
$\bss{h}^{(j)}$ is a $(2j)$-form, and in particular $\bss{h}^{(1)}=\bss{h}$.
Since $\bss{h}$ is non-degenerate, one has a
set of $n$ non-vanishing closed CKY tensors. In an even dimensional
spacetime  $\bss{h}^{(n)}$ is proportional to the totally antisymmetric
tensor whereas it is dual to a Killing vector in odd dimensions. In both cases 
such CKY tensor is trivial and can be excluded from the tower of hidden symmetries. The CKY tensors can be generated from the potentials $\bss{b}^{(j)}$
(cf. eq. \eqref{db})
\be
\bss{b}^{(j)}=\tens{b}\wedge \bss{h}^{\wedge (j-1)}\,,\quad
\bss{h}^{(j)}=\tens{d}\bss{b}^{(j)}\,.
\ee

Each $(2j)$-form $\bss{h}^{(j)}$ 
determines a $(D-2j)$-form of the Killing--Yano tensor
\be
\bss{f}^{(j)}=*\bss{h}^{(j)}\, .
\ee
In their turn, these tensors give rise to the Killing tensors
$\bss{K}^{(j)}$
\be\n{KK}
K^{(j)}_{ab}={1\over (D-2j-1)!(j!)^2} f^{(j)}_{\, \, \, \, \, a c_1\ldots c_{D-2j-1}}
f_{\, \, \, \, \, b}^{(j) \, \,  c_1\ldots c_{D-2j-1}}\, .
\ee
A choice of the coefficient in the definition \eq{KK} is 
adjusted to Section~V, cf. eq. \eqref{CKT}.
It is also convenient
to include the metric $\bss{g}$, which
is a trivial Killing tensor, as an element $\bss{K}^{(0)}$ of the
tower of the Killing tensors. The total number of irreducible 
elements of this {\em extended
tower} is $n$.\footnote{For example, in 5D spacetime where $n=2$,
this tower contains only one non-trivial
Killing tensor. For the 5D rotating black hole solution this Killing
tensor was first found in \cite{FSa,FSb} by using the Carter's method
of separation of variables in the Hamilton--Jacobi equation.}

\subsection{Canonical basis and canonical coordinates}

Similar to 4D case, let us consider the eigenvalue
problem for the conformal Killing tensor 
$H_{ab}=h_{ac}h_b^{\ \,c}$. It is easy to show that in the Euclidean domain
its eigenvalues $x^2$, 
\be\n{problem}
H^a_{\ b}v^{b}=x^2 v^a\, ,
\ee 
are real and non-negative. Using a modified Gram--Schmidt procedure it
is possible to show that there exists such an orthonormal basis in
which the operator $\tens{h}$ has the following structure:
\be\n{dar}
\mbox{diag}(0,\ldots,0,{\Lambda}_1,\ldots,{\Lambda}_p)\, ,
\ee
where ${\Lambda}_i$ are matrices of the form
\be
{\Lambda}_i=\left(   
\begin{array}{cc}
0 & -x_i {I}_i\\
x_i {I}_i& 0
\end{array}
\right)\, ,
\ee
and ${I}_i$ are unit matrices. Such a basis is known as the {\em
Darboux basis} (see, e.g., \cite{pras}). Its elements are unit
eigenvectors of the problem \eq{problem}.

For a non-degenerate 2-form $\bss{h}$ the number of zeros
in the Darboux decomposition \eq{dar} coincides with $\ve$. If all the
eigenvalues $x$ in \eq{problem} are different (we denote them
$x_{\mu}$, $\mu=1,\ldots,n$), the matrices
${\Lambda}_i$ are 2-dimensional. Denote the vectors of the Darboux
basis by $\bsn{e}{\mu}$ and $\bsnn{e}{\mu}\equiv \bsn{e}{n+\mu}$,
where $\mu=1,\ldots,n$. In an odd dimensional spacetime we also have
an additional basis vector $\bsn{e}{0}$ (the eigenvector of
\eq{problem} with $x=0$). Orthonormal vectors $\bsn{e}{\mu}$ and
$\bsnn{e}{\mu}$ span a 2-dimensional plane of eigenvectors of
\eq{problem} with the same eigenvalue $x_{\mu}$. We denote by
$\bse{\omega}{\mu}$ and $\bsee{\omega}{\mu}\equiv \bse{\omega}{n+\mu}$ (and
$\bse{\omega}{0}$ if $\ve=1$) the dual basis of 1-forms.
The metric $\tens{g}$ and the principal CKY tensor $\bss{h}$ in this basis take
the form
\ba
\tens{g}&=&\sum_{\mu=1}^n (\bse{\omega}{\mu}\bse{\omega}{\mu}+
\bsee{\omega}{\mu}\bsee{\omega}{\mu})+\ve \bse{\omega}{0}\bse{\omega}{0}\,
,\n{gab}\\
\bss{h}&=&\sum_{\mu=1}^n x_{\mu} \bse{\omega}{\mu}\wedge \bsee{\omega}{\mu}\,
.\n{hab}
\ea

\section{Hidden symmetries of higher--dimensional Kerr-NUT-(A)dS
spacetimes}

The most general known higher-dimensional solution describing rotating black
holes with NUT parameters in an asymptotically (Anti) de Sitter
spacetime (Kerr-NUT-(A)dS metric) was found by Chen, L\"u, and Pope \cite{pope}. In an (analytically continued) symmetric form it is written as \eq{gab} where
\be \n{bas_e}
\tens{\omega}^\mu=\frac{\tens{d}x_\mu}{\sqrt{Q_\mu}}\,,\quad
\tens{\omega}^{\bar{\mu}}=\sqrt{Q_\mu}\sum_{k=0}^{n-1}A^{(k)}_{\mu} \tens{d}\psi_k\,,
\quad
\tens{\omega}^{0} = (c/A^{(n)})^{1/2}
\sum_{k=0}^nA^{(k)}\tens{d}\psi_k\, .
\ee
Here 
\begin{gather}
Q_{\mu}=X_{\mu}/U_{\mu}\,,\quad
U_{\mu}=\prod_{\nu\ne\mu}(x_{\nu}^2-x_{\mu}^2)\,,\nonumber\\
A_{\mu}^{(k)}=\!\!\!\!\!\sum_{\substack{\nu_1<\dots<\nu_k\\\nu_i\ne\mu}}\!\!\!\!\!x^2_{\nu_1}\dots x^2_{\nu_k},\quad 
A^{(k)}=\!\!\!\!\!\sum_{\nu_1<\dots<\nu_k}\!\!\!\!\!x^2_{\nu_1}\dots x^2_{\nu_k}\;\label{vztahy}.
\end{gather}
Metric coefficients $X_{\mu}$ are functions of $x_{\mu}$ only, 
and for the Kerr-NUT-(A)dS solution take the form 
\be\n{XX}
X_{\mu}=\sum\limits_{k=\varepsilon}^{n}c_kx_{\mu}^{2k}-2b_{\mu}x_{\mu}^{1-\varepsilon}-\frac{\varepsilon c}{x_{\mu}^2}\,.
\ee 
Time is denoted by $\psi_0$, azimuthal coordinates by $\psi_k$,
${k=1,\dots,m}$, and ${x_\mu}$, ${\mu=1,\dots,n}$, stand for  
`radial' and latitude coordinates. Here we have introduced $m=n-1+\ve$. The physical metric with proper signature is recovered when standard radial coordinate $r=-ix_n$ and new parameter $M=(-i)^{1+\epsilon}b_n$ are introduced. 
The total number of constants which enter the solution is $2n+1$: $\ve$ constants $c$, $n+1-\ve$ constants $c_k$ and $n$ constants
$b_{\mu}$. The form of the metric is invariant under a 1-parameter 
scaling coordinate transformations, thus a total number of
independent parameters is $D-\ve$. 
These parameters are related to the cosmological constant, mass,
angular momenta, and NUT parameters. One of them may be used to
define a scale, while the other $D-1-\ve$ parameters can be made
dimensionless. (For more details see \cite{pope}.) Similar to the 4D
case, the signature of the symmetric form of the metric depends on
the domain of $x_\mu$'s and the signatures of $X_\mu$'s.

Hamamoto, Houri, Oota, and Yasui \cite{hoy3}
derived explicit formulas for the curvature and demonstrated that in all 
dimensions this metric obeys the Einstein equations,
\be\n{Eeq}
R_{ab}=\Lambda g_{ab}\,,\quad 
\Lambda=(D-1)(-1)^n c_n\, .
\ee
The limit of flat spacetime is recovered when $c_n=0$ and all of the parameters $b_{\mu}$ are zero (equal one to each other) in the even (odd) dimensional case.  The metric belongs to the class of special algebraic type {\bf D} solutions \cite{hoy3} of the higher-dimensional algebraic classification  \cite{classification}. It 
may be understood as a higher-dimensional
generalization of the 4-dimensional Ker-NUT-(A)dS solution obtained by
Carter \cite{Car:68c}. 
Moreover, the coordinates $(x_\mu,\psi_k)$ used in the metric
are the higher-dimensional analogue of the canonical coordinates 
\cite{Car:68c, CDP}.

It was shown in \cite{KF,FK} that this spacetime possesses a principal
CKY tensor $\bss{h}$ which has the form \eq{hab} and its potential
$\bss{b}$, $\tens{h}=\tens{db}$, is\footnote{In fact, this potential
generates a principal CKY tensor for a general form of the metric
\eq{gab},\eq{bas_e}, and \eqref{vztahy} with arbitrary functions
$X_{\mu}(x_{\mu})$.}
\be\n{bbbb}
\bss{b}={1\over 2}\sum_{k=0}^{n-1} A^{(k+1)} \tens{d}\psi_k\, .
\ee
The tower of Killing tensors \eqref{KK} associated with this principal CKY tensor is 
\cite{KKPF}
\be
\tens{K}^{(j)}=\sum_{\mu=1}^n A_{\mu}^{(j)} 
(\bse{\omega}{\mu}\bse{\omega}{\mu}+
\bsee{\omega}{\mu}\bsee{\omega}{\mu})
+\ve A^{(j)}\bse{\omega}{0}\bse{\omega}{0}\,,\qquad
j=1,\dots,n-1\, .
\ee

In the Kerr-NUT-(A)dS spacetime besides this tower of Killing tensors
also all the isometries follow  from  the existence of the principal CKY tensor $\tens{h}$ \cite{KKPF}. 
The {\em primary Killing vector} $\bss{\xi}^{(0)}$ is given by \eq{prime} and reads
\begin{equation}
\bss{\xi}^{(0)}=\pa_{\psi_0}\,.
\end{equation} 
This vector plays a special role. After the analytical continuation to the physical space it is the only one timelike in the black hole exterior.
Besides the primary Killing vector, 
this spacetime has $n-1$ {\em secondary Killing vectors}  $\bss{\xi}^{(j)}$,
\be\n{kvs}
\xi^{(j)\,a}=K^{(j)}\,\!^a_{\ b}\xi^b=(\partial_{\psi_j})^a\,,\qquad
j=1,\dots,n-1\, .
\ee
In odd dimensions the last Killing vector is given by
the ${n}$-th Killing--Yano tensor  $\bss{f}^{(n)}$, which in the
Kerr-NUT-(A)dS spacetime  turns out to be 
\begin{equation}
\tens{\xi}^{(n)}=\bss{f}^{(n)}=\pa_{\psi_n}\,.
\end{equation}

The total number of these  Killing vectors is $n+\ve$. For a geodesic
motion they give $n+\ve$ linear in momentum integrals of motion, 
$\Psi_j=\xi^{(j)}_{\ a} p^a$. The
extended tower of the Killing tensors $\bss{K}^{(j)}$ ($j=0,\ldots n-1$) gives
$n$ additional integrals of motion, which are quadratic in the
momentum, $\kappa_j=K^{(j)}_{\ \, ab}p^ap^b$. Thus the total number of conserved quantities for a geodesic
motion is $2n+\ve$, that is it coincides with the number of 
spacetime dimensions $D$. It is possible to show that these integrals of
motion are independent and in involution, so that the geodesic motion in the
Kerr-NUT-(A)dS spacetime is completely integrable
\cite{PKVK,KKPV}.
The components of the particle momentum $\tens{p}$ can be written as functions 
of the integrals of motion as follows \cite{KKPF}:
\begin{equation}\label{basisvectors}
\begin{aligned}
p_\mu =&\, \frac{\sigma_\mu}{(X_{\mu}U_\mu)^{1/2}}\Bigl[\,
X_\mu \sum_{j=0}^{m}(-x_{\mu}^2)^{n-1-j}\kappa_j
-\bigl(\sum_{j=0}^{m}(-x_{\mu}^2)^{n-1-j}\Psi_j\bigr)^2\Bigr]^{1/2}\,,\\
p_{\bar\mu}=&\, \frac{1}{(X_\mu U_\mu)^{1/2}}\,
\sum_{j=0}^{m}(-x_{\mu}^2)^{n-1-j}\Psi_j\,\,,\\
p_{0} =&\, \bigl(cA^{(n)}\bigr)^{-1/2}\Psi_n\,,\quad 
\kappa_n=\frac{\Psi_n^2}{c}\,.
\end{aligned}
\end{equation} 
Constants $\sigma_\mu$ which denote the choice of signs 
are independent of one another.

It should be emphasized that, similar to 4D case, the coordinates $(x_\mu,\psi_k$) in the metric \eq{gab}, \eq{bas_e}, \eq{vztahy} have a well defined geometrical meaning. The
`essential' coordinates $x_{\mu}$ are connected with the eigenvalues of
the principal CKY tensor $\bss{h}$ (see \eq{hab}), while the Killing
coordinates $\psi_{k}$ are defined by the Killing vectors generated by
the principal CKY tensor. It is this invariant definition of the
coordinates what makes  this form of the metric so convenient for calculations.

The existence of a principal CKY tensor imposes non-trivial
restrictions on the geometry of the spacetime. Namely, the following
result was  proved in~\cite{hoy1, hoy2}. Let $\bss{h}$ be a principal CKY
tensor and $\bss{{\xi}^{(0)}}$ be its primary Killing vector. Then if
\be\n{cond}
{\cal L}_{{\xi}^{(0)}} \bss{h}=0\, ,
\ee
the only solution of the higher-dimensional Einstein equations with the cosmological
constant \eq{Eeq} is the Kerr-NUT-(A)dS spacetime. (Here ${\cal L}_{u}$ is a Lie
derivative along the vector $\bss{u}$.)

Let us finally mention that recently it was shown \cite{sekr} that the following
operators (cf. eq. \eqref{opK}):
\ba\n{LLL}
{\hat \xi}_{(k)}&=&-i\xi^{(k)a}\partial_{a}\,,\quad k=0,\ldots,m\, ,\\
{\hat K}_{(j)}&=&-{1\over\sqrt{|g|}}\,
\partial_{a}\Bigl(\sqrt{|g|}K^{(j)\,\! a  b}\partial_{b}\Bigr)\,,\quad j=0,\ldots,n-1\, ,
\n{KKK}
\ea
determined by a principal CKY tensor, form a complete set of commuting operators 
for the Klein--Gordon equation in the Kerr-NUT-(A)dS background.

\section{Hidden symmetries and separation of variables}
The massive scalar field equation
\be\n{seq}
\Box \Phi-\mu^2\Phi=0\, ,
\ee
in the Kerr-NUT-(A)dS metric allows a complete separation of variables
\cite{FKK}. Namely, the solution can be decomposed into modes
\be\n{mod}
\Phi=\prod_{\mu=1}^n R_{\mu}(x_{\mu})\prod_{k=0}^{m}e^{i\Psi_k
\psi_k}\, .
\ee
Substitution of \eq{mod} into equation \eq{seq} results in the
following second order ordinary differential equations for functions
$R_{\mu}(x_{\mu})$
\be\label{solution}
(X_{\mu}R'_{\mu})'+\ve{X_{\mu}\over
x_{\mu}}R'_{\mu}+\left(V_{\mu}-{W_{\mu}^2\over X_{\mu}}\right)R_{\mu}=0\, .
\ee
Here
\be\n{fun}
W_{\mu}=\sum_{k=0}^{m}\Psi_k(-x_{\mu}^2)^{n-1-k}\,,\quad
V_{\mu}=\sum_{k=0}^{m}\kappa_k(-x_{\mu}^2)^{n-1-k}\, .
\ee
Here $\kappa_0=-\mu^2$
and for $\ve=1$ we put $\kappa_n=\Psi_n^2/c$. The parameters
$\kappa_k$ ($k=1,\ldots n+\ve-1$)
are separation constants. Using \eq{KKK} one has ${\hat K}_{(0)}=-\Box$.
Since all the operators \eq{LLL}--\eq{KKK} commute with one another,
their common eigenvalues can be used to specify the modes. It is possible
to show \cite{sekr} that the eigenvectors of these commuting operators are
the modes \eq{mod} and one has
\be
{\hat \xi}_{(k)}\Phi=\Psi_k\Phi\,,\quad 
{\hat K}_{(j)}\Phi=\kappa_j \Phi\, .
\ee
Similar to the case discussed in 4D, in the symmetric form of the metric \eq{gab}, \eq{bas_e}, \eq{vztahy} all the equations \eqref{solution} `look the same'. 
However, after the transformation to the physical space the equation for $R_n$ plays the role of an equation for radial modes, whereas the other equations present the eigenvalue problem. For a discussion of special sub-cases of these equations see, e.g, 
\cite{Cardoso} and reference therein.   

The Hamilton--Jacobi equation for geodesic motion 
\be
{\partial S\over \partial \lambda}+g^{ab}\partial_a S \partial_b S=0\,,
\ee
in the Kerr-NUT-(A)dS spacetime also allows a complete separation of
variables \cite{FKK}
\be
S=\mu^2\lambda+\sum_{k=0}^{m}\Psi_k\psi_k+\sum_{\mu=1}^n
S_{\mu}(x_{\mu})\, .
\ee
The functions $S_{\mu}$ obey the first order ordinary differential
equations
\be
{S'_{\mu}}^2={V_{\mu}\over X_{\mu}}-{W_{\mu}^2\over X_{\mu}^2}\, ,
\ee
where the functions $V_{\mu}$ and $W_{\mu}$ are defined in \eq{fun}.

Recently it was shown that the massive Dirac equation in the
Kerr-NUT-(A)dS spacetime also allows the separation of variables
\cite{oy}. It was also proved that the Nambu-Goto equations 
for a stationary test string
in the  Kerr-NUT-(A)dS background are completely integrable
\cite{KF_string} .

\section{Conclusions}

In this review we  discussed recent developments of the theory of
higher-dimensional black holes. We focused mainly on the problem of
hidden symmetries and separation of variables. (For more general
discussion of the modern status of the theory of higher-dimensional
black holes see a recent review \cite{Em:2008}.) In our presentation,
we started with a description of known results concerning
four-dimensional black holes. We found this important since many of
the properties of higher-dimensional isolated black holes are quite
similar to the properties of black holes in four dimensions. What we
tried to illustrate in this presentation is that most of the
important properties of a stationary isolated black hole solution
follow from the existence of what is called a {\em principal
conformal Killing--Yano tensor}. This is true in four dimensions and,
as recent studies demonstrated, in higher dimensions as well. 

The Kerr-NUT-(A)dS metric is the most general known solution
describing the higher-dimensional rotating black hole spacetime with
NUT parameters in an asymptotically (Anti) de Sitter background.  It
possesses a principal CKY tensor $\bss{h}$ which determines the
hidden symmetries of this spacetime. The 2-form $\bss{h}$ generates
a tower of Killing--Yano and Killing tensors, which allow complete
integrability of geodesic equations and separability of the
Hamilton--Jacobi, Klein--Gordon, and Dirac equations. Moreover, if
the higher-dimensional solution of the Einstein equations with the
cosmological constant  (often called the {\em Einstein space}) allows
a principal CKY tensor obeying \eq{cond}, it coincides with the
Kerr-NUT-(A)dS metric \cite{hoy3}. These remarkable properties of
higher-dimensional rotating black holes resemble the well known
miraculous properties of the Kerr spacetime described partly in
the introduction.  In four dimensions all of the Einstein spaces which
possesses the KY tensor are of the Petrov type {\bf D}. A natural conjecture in
higher dimensions is that any Einstein space which possesses the
principal CKY tensor is  of the special algebraic type {\bf D} of
higher-dimensional classification \cite{classification}.

Does this analogy go further? We focused on black holes, but in the
higher-dimensional gravity there exists a variety of other black
objects such as black rings or black saturns. Do these spacetimes
also possess hidden symmetries? Such symmetries would be very helpful
for the study of the stability of these `exotic' objects. 
However, for example black rings  are of the algebraic type ${\bf
I}_i$ \cite{PrPr} and thus their symmetry properties might be quite
different from the properties of higher-dimensional black holes.
In particular, if the above conjecture is correct they do not allow a
principle CKY tensor and so nor a corresponding tower of hidden symmetries. 

The results on the separability till now obtained  can be used for
the study of the particle and light propagation in higher-dimensional
rotating black hole spacetimes. They allow one to calculate the
contribution of the scalar and Dirac fields to the bulk Hawking
radiation of higher-dimensional rotating black holes, without any
restrictions on black hole parameters.

An important open question is a separability problem for
the gravitational perturbations in higher-dimensional
black hole spacetimes.  A certain progress in this direction was achieved
recently (see, e.g., \cite{pr}, \cite{Murata}).
These results are very important for the study of the stability of such
black holes and different aspect of the Hawking radiation produced by
them. Another important direction of research is the study of the
quasinormal modes in higher dimensions.
Most of the results obtained in these directions (see, e.g., \cite{n} and
references therein) assumed some additional restrictions 
on the parameters characterizing black hole solutions. 
This remind a situation for the Klein--Gordon and
Dirac equations before the general results on their separability were
proved.

An important open question is:  Can the higher spin massless field
equations be decoupled in the background of the general Kerr-NUT-(A)dS metric
and do they allow separation of variables? Recent result of \cite{oy}
on the separability of the massive Dirac equation is quite promising.
Separability of the higher spin equations, and especially the
equations for the gravitational perturbations, would provide one with
powerful tools important, for example, for the study of the stability
of higher-dimensional black hole solutions. One might  hope that it
will not take too long time before this and other interesting open
questions connected with the existence of hidden symmetries in
higher-dimensional black holes will find their answers.

\section*{Acknowledgements}

This paper is an extended version of the talk given by one of the
authors, V.F., at the ``Peyresq Physics 12'' meeting on ``Micro and
Macro structure of spacetime'' held in Peyresq in June 18-24, 2007. He
appreciate very much stimulating scientific atmosphere of Peyresq
village created by Mady Smets, who devoted this place to humanistic
activities.  V.F thanks  ``OLAM Association for Fundamental Research,
Brussels, Belgium'' for its support, and  Diane and Edgard Gunzig for
their hospitality. V.F. is also grateful to the Natural Sciences and
Engineering Research Council of Canada and the Killam Trust for
partial support. D.K. acknowledges the Golden Bell Jar Graduate
Scholarship in Physics at the University of Alberta.

\end{document}